\documentclass[10pt,superscriptaddress,twocolumn,amsmath,amssymb,aps,prb]{revtex4}
\usepackage{mathrsfs}
\usepackage{graphicx}
\usepackage{dcolumn}
\usepackage{bm}

\newcommand{\ucite}[1]{\scalebox{1.4}[1.4]{\raisebox{-0.8ex}{\cite{#1}}}}
\newcommand{\be}{\begin{equation}}
\newcommand{\ee}{\end{equation}}
\newcommand{\bea}{\begin{eqnarray}}
\newcommand{\eea}{\end{eqnarray}}
\newcommand{\nn}{\nonumber}
\begin{document}

\title{ Robust d-wave pairing symmetry in  multi-orbital  cobalt high temperature superconductors}
\author{Yinxiang Li}\thanks{these authors contributed equally to this work }
\affiliation{Beijing National Laboratory for Condensed Matter Physics,
and Institute of Physics, Chinese Academy of Sciences, Beijing 100190, China}
\author{Xinloong Han}\thanks{these authors contributed equally to this work }
\affiliation{Beijing National Laboratory for Condensed Matter Physics,
and Institute of Physics, Chinese Academy of Sciences, Beijing 100190, China}
\author{Shengshan Qin}
\affiliation{Beijing National Laboratory for Condensed Matter Physics,
and Institute of Physics, Chinese Academy of Sciences, Beijing 100190, China}
\author{Congcong Le}
\affiliation{Beijing National Laboratory for Condensed Matter Physics,
and Institute of Physics, Chinese Academy of Sciences, Beijing 100190, China}
\author{Qiang-Hua Wang}
\affiliation{National Laboratory of Solid State Microstructures and School of Physics, Nanjing University, Nanjing, 210093, China}
\affiliation{Collaborative Innovation Center of Advanced Microstructures, Nanjing University, Nanjing, 210093, China}
\author{Jiangping Hu}\email{jphu@iphy.ac.cn}
\affiliation{Beijing National Laboratory for Condensed Matter Physics,
and Institute of Physics, Chinese Academy of Sciences, Beijing 100190, China}
\affiliation{Collaborative Innovation Center of Quantum Matter,
Beijing, China}
\affiliation{Kavli Institute of Theoretical Sciences, University of Chinese Academy of Sciences,
Beijing, 100190, China}
\date{\today}

\begin{abstract}
The pairing symmetry of the newly proposed cobalt high temperature (high-$T_c$)  superconductors formed by vertex shared cation-anion tetrahedral complexes is studied by the methods of mean field, random phase approximation (RPA) and functional renormalization group (FRG) analysis. The results of all these methods show that the  $d_{x^2-y^2}$ pairing symmetry is robustly favored near half filling.  The RPA and FRG methods, which are  valid in weak interaction regions,  predict that  the superconducting state is also strongly orbital selective, namely the $d_{x^2-y^2}$ orbital that has the largest density near half filling among the three $t_{2g}$ orbitals dominates superconducting pairing.  These results suggest that the new materials, if synthesized, can provide indisputable test to high-$T_c$ pairing mechanism and the validity of different theoretical methods.
\end{abstract}


\maketitle
\section{Introduction}\label{sectioni}
Since the discovery of high temperature (high-$T_c$) superconductors, cuprates \cite{Cu} and iron-based superconductors \cite{Fe}, searching for new high $T_c$ superconductors and explaining their pairing mechanism become  crucial tasks  in condensed matter physics.  However, as the pairing mechanism still remains illusive, few useful theoretical clues have been provided to predict or identify new high-$T_c$ materials.

Recently, through extensive analysis of theoretical and experiment results, some of us have suggested that   the two known high $T_c$ families share a common electronic property--those $d$ orbitals with the strongest in-plane $d-p$ coupling in the cation-anion complex are isolated near Fermi energy \cite{jpHuprx,jpHusciencebult,hu&yuan}. From the viewpoint of magnetically-driven superconducting mechanism, this property implies that the effective antiferromagnetic (AFM) superexchange interaction is  the underlining driven force of high $T_c$ superconductivity. Moreover, this crucial property is largely absent in other correlated electronic systems. Thus, the uniqueness of this electronic structure  suggests that it can be the key to  predict or identify possible high $T_c$ materials.  Finding a new family of high $T_c$ materials with the common electronic structure  can provide a convincing test to superconducting pairing mechanism.

Following the above theoretical guide, two explicit proposals have been made for Co/Ni transition metal compounds as  potential candidates for high $T_c$ superconductors \cite{jpHuprx,ZnCo}. The first one includes a two dimensional hexagonal lattice formed by edge-shared trigonal biprymidal complexes \cite{jpHuprx} and the second one hosts a two dimensional square lattice formed by vertex-shared tetrahedra complexes \cite{ZnCo}. In both cases, a $d^7$ filling configuration meets the required electronic condition to be a potential high $T_c$ material.

The second proposal is particularly interesting because it provides an explicit playground to bridge and unify iron-based superconductors and cuprates. On  the one hand, the square lattice formed by vertex-shared tetrahedra complexes is similar to the one formed by vertex-shared octahedra complexes in cuprates.  Based on the empirical Hu-Ding principle \cite{Hu-Ding} of the pairing symmetry selection based on the matching between the pairing form factors from the AFM superexchange interactions and Fermi surfaces,  we can easily argue that a superconducting state with a d-wave pairing symmetry is favored. On the other hand,  similar to those of iron-based superconductors, the electronic structure is attributed to the multi $t_{2g}$ orbitals. Those antiferromagnetic superexchange couplings, similar to those iron-based superconductors that are responsible for superconductivity,  are maintained.  Although the proposal structure has not been synthesized experimentally,  the square CoA$_2$ (A=S,Se) lattice can be theoretically constructed through the well known cubic Zinc-Blende structure.  For example,  using MBE (Molecular beam epitaxy),   we may grow ZnCoS$_2$ \cite{ZnCo} which hosts layered square lattices with vertex-shared CoS$_4$ tetrahedra by replacing half of Zn atoms in ZnS by Co \cite{ZnCo}.

In this paper, we carry  out a systematic investigation of the superconducting state in this family of materials  under a variety of theoretical methods, including the mean field approach, random phase approximation (RPA) and functional renormalization groups (FRG).  All these methods show that the d-wave pairing symmetry is robustly favored in this system while the  pairing strength among different $t_{2g}$ orbitals depends on theoretical methods. In particular,  as the RPA and FRG methods are very sensitive to the density of states near Fermi energy,  both methods  predict that  the superconducting state is strongly orbital selective. The $d_{x^2-y^2}$ orbital that has the largest density near half filling among the three t$_{2g}$ orbitals  dominates superconducting pairing within the reasonable range of the interaction parameters.

The paper is organized as follows. In Sec. \ref{sectionii}, we present the mean field analysis of the effective $t-J$ model to show the pairing symmetry near the half filling regime. We find that the $d_{x^2-y^2}$ wave pairing from $d_{xz}$ and $d_{yz}$ orbitals is competing with that from $d_{x^2-y^2}$ orbital. In Sec. \ref{sectioniii}, we provide the gap function within three-orbital Hubbard model by RPA method. The symmetry of $d_{x^2-y^2}$ wave is always the largest pairing strength which mainly from the contribution of $d_{x^2-y^2}$ orbital electrons. In Sec. \ref{sectioniv}, we find the full pockets have $d$-wave pairing symmetry by FRG analysis. The pairing form factor mainly situates at the $\beta$ pocket, showing a very strong $d_{x^2-y^2}$ orbital selection when the system is doped near the half filling.  Finally, we summarize and discuss these results  in Sec. \ref{sectionv}.
\section{The mean field results from the multi-orbital t-J model}\label{sectionii}
We use the  model that is derived for the single layer CoS$_2$ in  the ZnCoS$_2$ compound \cite{ZnCo}. Since only the three $t_{2g}$ orbitals appear near the Fermi level and the nearest neighbour (NN) antiferromagnetic (AFM) interaction is  responsible for superconductivity (SC) \cite{ZnCo}, the following calculations are all done based on the three bands model in ref. \ucite{ZnCo} with the NN AFM interaction $J$. The tight-binding model  is given in ref. \ucite{ZnCo} as
\begin{equation}
\emph{H}_{0}=\left(
\begin{array}{ccccc}
H_{11}-\mu & H_{12} & H_{13}\\
H_{21} & H_{22}-\mu & H_{23}\\
H_{31} & H_{32} & H_{33}-\mu\\
\end{array}
\right).
\end{equation}
The elements of $H_0$ matrix are given by \cite{ZnCo}
\begin{eqnarray}
\label{eq2}
H_{11} & = & \epsilon_{1}+2t^{11}_{x}cos(k_x) +2t^{11}_{y}cos(k_y)+4t^{11}_{xy}cos(k_x)cos(k_y)\nonumber
\\
&& +2t^{11}_{xx}cos(2k_x) +2t^{11}_{yy}cos(2k_y),\nonumber \\
H_{12} & = & -4t^{12}_{xy}sin(k_x)sin(k_y)\nonumber
\\
H_{13} & = & 2it^{13}_xsin(k_x)+4it^{13}_{xy}sin(k_x)cos(k_y)+2it^{13}_{xx}sin(2k_x)\nonumber
\\
H_{22} & = & \epsilon_{2}+2t^{22}_{x}cos(k_x) +2t^{22}_{y}cos(k_y)+4t^{22}_{xy}cos(k_x)cos(k_y)\nonumber
\\
&& +2t^{22}_{xx}cos(2k_x) +2t^{22}_{yy}cos(2k_y),\nonumber
\\
H_{23} & = & 2it^{23}_ysin(k_y)+4it^{23}_{xy}sin(k_y)cos(k_x)+2it^{23}_{xx}sin(2k_y)
\nonumber
\\
H_{33} & = & \epsilon_{3}+2t^{33}_{x}(cos(k_x)+cos(k_y))+4t^{33}_{xy}cos(k_x)cos(k_y)\nonumber
\\
&& +2t^{33}_{xx}(cos(2k_x)+cos(2k_y)).
\end{eqnarray}
Without a further specification, we take all energy parameters in the unit of eV. In above equation, $\epsilon_{1}=\epsilon_{2}=3.7314$ and $\epsilon_{3}=4.1241$ are the onset energy of $d_{xz,yz}$ and $d_{x^2-y^2}$. The corresponding hopping parameters are $t^{11}_{x}=t^{22}_{y}=0.4391$, $t^{11}_{y}=t^{22}_{x}=0.1408$, $t^{11}_{xy}=t^{22}_{xy}=-0.0162$, $t^{12}_{xy}=0.021$, $t^{13}_x=t^{23}_y=0.0057$,
$t^{13}_{xy}=t^{23}_{xy}=-0.0061$, $t^{33}_x=0.1824$,$t^{33}_{xy}=0.011$, $t^{11}_{xx}=t^{22}_{yy}=0.0688$,
$t^{11}_{yy}=t^{22}_{xx} {=}-0.0025$, $t^{13}_{xx}=t^{23}_{yy}=0.0107$, $t^{33}_{xx}=-0.0299$, in which x(y) labels the hopping between two NN sites along x (y) directions, xy labels the hopping between two next NN sites, and xx (yy) labels the hopping between two third NN sites along x (y) directions. The Fermi surface (FS) of three-orbital model is plotted in Fig. 1(a) with 0.5 electron doped per site away from half filling. The Fermi surface includes two hole pockets around the $\Gamma$ point and one electron pocket around the $M$ point in the first Brillouin zone (BZ).  In the spirit of the $t-J$ model, the kinetic energy is also subject to a full suppression of on-site double occupancy. In the mean field calculation, we can absorb this suppression into an effective overall renormalization factor of the bare band structure \cite{Kotliar, seohu}. In general, this renormalization factor is also  doping dependent. As here we focus on qualitatively obtaining the pairing symmetry,   we can simply stick to the bare band structure by rescaling the interaction parameters.
 \begin{figure}
\centerline{\includegraphics[width=0.5\textwidth]{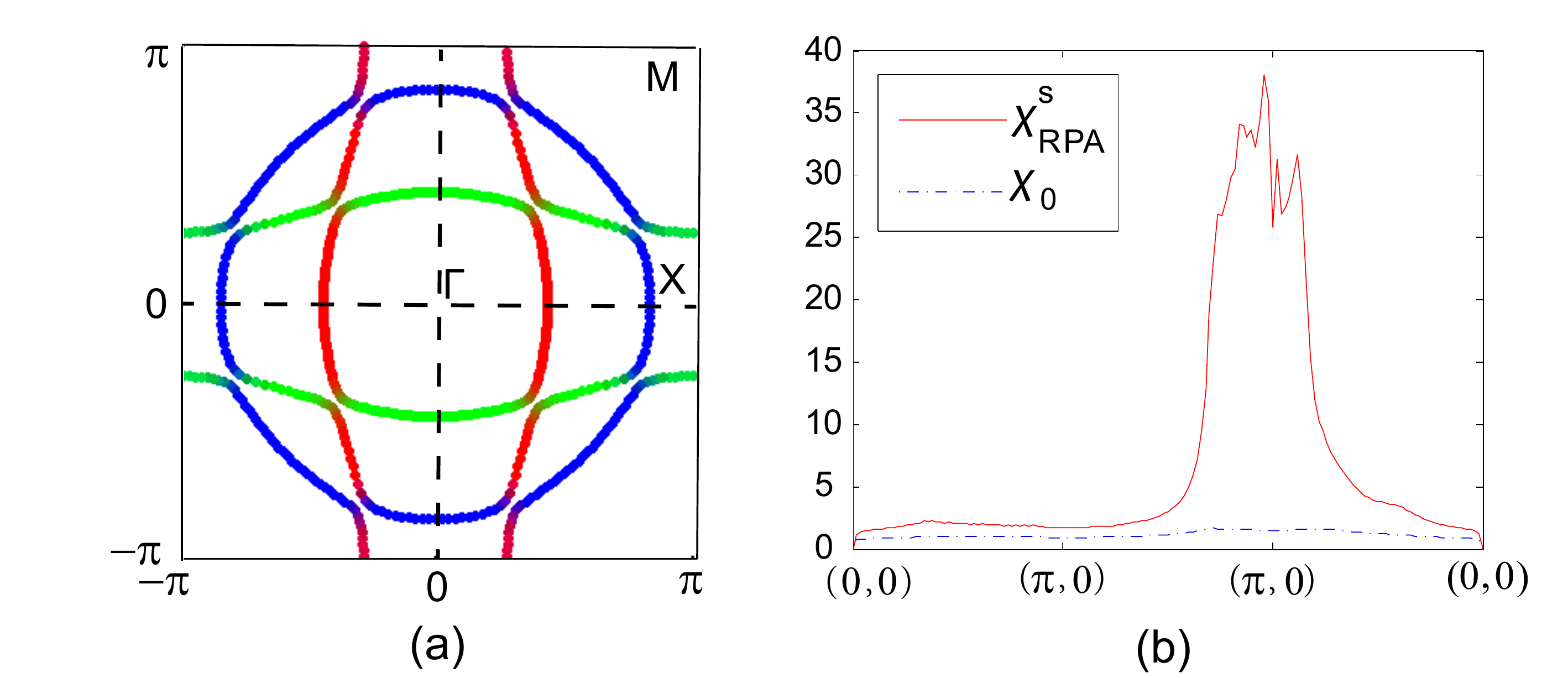}}
\caption{(color online) The orbital contributions of the different FS sheets are shown color coded:$d_{xz}$(red), $d_{yz}$(green) and $d_{x^{2}-y^{2}}$(blue) at the  0.5 electron doping with respect to the half filling  in (a).The bare susceptibility and RPA spin susceptibility  with $U=0.48$ and $J_{H}/U=0.2$ in (b).}
\end{figure}

Now, we consider the effective AFM interactions in this family.  The effective AFM interactions are generated by the  superexchange  process. Moreover,  the intra-orbital AFM couplings  dominate over the inter-orbital AFM couplings.  The intra-orbital AFM couplings can be estimated as
 \begin{equation}
\begin{aligned}
J_{\alpha}=4t^{2}_{eff,\alpha}(\frac{1}{U_{d}}+\frac{1}{U_{d}+\Delta_{pd}}),
\end{aligned}
\end{equation}
where $U_d$ is the Coulomb interaction for the $d_{\alpha}$ orbital, $t_{eff,\alpha}=t^{2}_{pd,\alpha}/(U_{d}+\Delta_{pd})$ is the effective intra-orbital hopping between the $d_{\alpha}$ orbitals at two NN sites and $\Delta_{pd}=\varepsilon_{d}-\varepsilon_{p}$ is the energy difference between the $d_{\alpha}$ orbital and $p$ orbital.   The hopping parameters and onsite energy for $p$ and $d_{\alpha}$ orbitals can be obtained by DFT calculation \cite{ZnCo} and the Coulomb interaction for the $d$ orbitals $U_d$ is set to be $3.0eV$ \cite{ZnCo}. After detailed calculations from DFT results, we found that for the $d_{x^2-y^2}$ orbital, $J_{x^2-y^2}\sim 0.07$. For the $d_{xz}$ and $d_{yz}$ orbitals, the couplings are different along the two different direction so that we denote them as $J_{xz}^{x}=J_{yz}^{y}\sim 0.25$ and $J_{xz}^{y}=J_{yz}^{x}\sim 0.03$. In the following, we fix the ratio between different $J_{\alpha}$ according to these estimation values in our calculation and vary  them by multiplying a single interaction scaling parameter $J$. The value of $J$ can be viewed as the renormalization factor of the bare band from the electron-electron correlation.

We can use the standard mean field method to decouple the AFM interaction in the superconducting channel. The superconducting order parameters in the spin singlet pairing channel are defined as
\begin{eqnarray}
\Delta_{<rr'>,\alpha}=J_{\alpha}<c_{\alpha,r,\uparrow}c_{\alpha,r^{'},\downarrow}-c_{\alpha,r,\downarrow}c_{\alpha,r^{'},\uparrow})>,
\end{eqnarray}
where $<rr'>$ represents two NN sites and $c_{\alpha,r,\sigma}$ are Fermionic operators.
In a uniform superconducting state,  the superconducting order parameters are also translation invariant.  Considering the three $t_{2g}$ orbitals, we have six independent pairing order parameters, $ \Delta_{\alpha}^{a}$ with $a$ denoting the directions and $\alpha$ denoting orbitals.  Since the lattice equivalently has  the $D_{4h}$ symmetry by a gauge transformation,  the order parameters can form two one-dimensional irreducible representations, namely, a s-wave pairing ($A_{1g}$ ) state and a d-wave pairing ($B_{1g}$) state, as follows,
\begin{equation}
\begin{aligned}
\Delta^{s}_{1}=&(\Delta_{xz}^{x}+\Delta_{yz}^{y})/2,\\
\Delta^{d}_{1}=&(\Delta_{xz}^{x}-\Delta_{yz}^{y})/2,\\
\Delta^{s}_{2}=&(\Delta_{xz}^{y}+\Delta_{yz}^{x})/2,\\
\Delta^{d}_{2}=&(\Delta_{xz}^{y}-\Delta_{yz}^{x})/2,\\
\Delta^{s}_{3}=&(\Delta_{x^{2}-y^{2}}^{x}+\Delta_{x^{2}-y^{2}}^{y})/2,\\
\Delta^{d}_{3}=&(\Delta_{x^{2}-y^{2}}^{x}-\Delta_{x^{2}-y^{2}}^{y})/2.\\
\end{aligned}
\end{equation}
Among these six order parameters, $\Delta^{s,d}_2$ are much smaller than the others because the $d-p$ couplings for the $d_{xz,yz}$ orbitals along the corresponding directions are very small.  Therefore, we ignore $\Delta^{s,d}_2$ in the following calculations.
In the momentum space, the mean-field Hamiltonian can be written as
\begin{equation}
\emph{H}=\left(
\begin{array}{cc}
H_{0}(k) & \Delta(k)\\
\Delta^{\dagger}(k) & -H^{\ast}_{0}(-k)\\
\end{array}
\right),
\end{equation}
where
\begin{equation}
\Delta(k)=\left(
\begin{array}{ccc}
\Delta_{11}(k) & 0 & 0\\
0 & \Delta_{22}(k) & 0\\
0 & 0 & \Delta_{33}(k)\\
\end{array}
\right),
\end{equation}
and
\begin{equation}
\begin{aligned}
\Delta_{11}(k)=&(\Delta^{s}_{1}+\Delta^{d}_{1})cos(k_{x})\\
\Delta_{22}(k)=&(\Delta^{s}_{1}-\Delta^{d}_{1})cos(k_{y})\\
\Delta_{33}(k)=&\Delta^{s}_{3}(cos(k_{x})+cos(k_{y}))\\
               &+\Delta^{d}_{3}(cos(k_{x})-cos(k_{y})).
\end{aligned}
\end{equation}

\begin{figure}
\centerline{\includegraphics[width=0.5\textwidth]{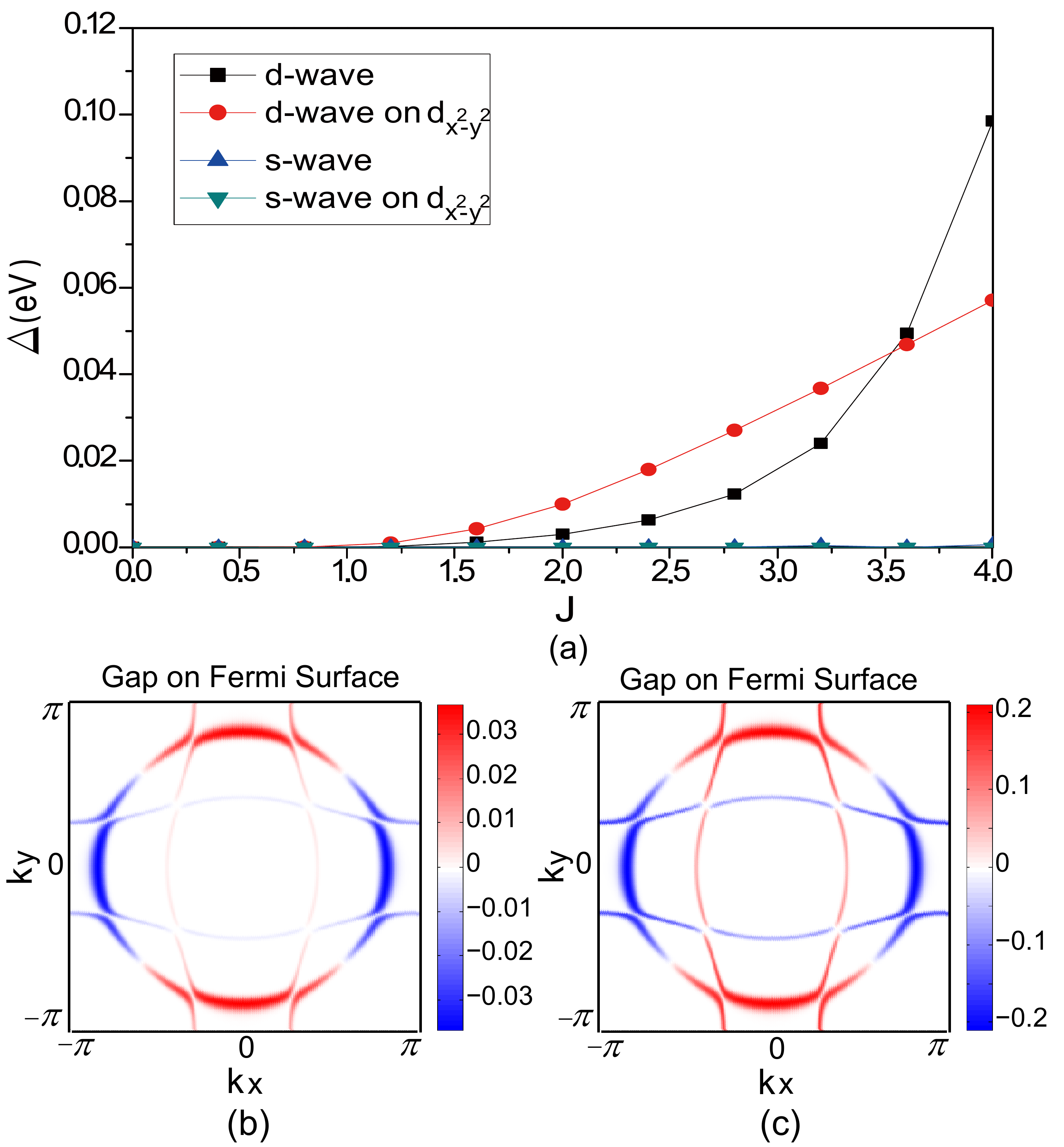}}
\caption{(color online) The mean field phase diagram with NN antiferromagnetic parameter $J$ in (a). The $d_{x^{2}-y^{2}}$-wave pairing symmetry with $J=2.0$ in (b) and $J=4.0$ in (c).}
\end{figure}

By fixing the ratio between different AFM interactions,  we can calculate the mean-field phase diagram for the pairing symmetry with respect to a single interaction scaling parameter $J$. The result is shown in the Fig. 2(a). The most favored pairing symmetry is the $d$ wave within all parameter ranges. The contribution from the $d_{xz}$ and $d_{yz}$ orbitals increases significantly when $J$ increases. When $J$ is less than 3.5, the $d$-wave pairing on the $d_{x^{2}-y^{2}}$ orbital which has the geometric factor $cos(k_x)-cos(k_y)$  is stronger than those on the other two orbitals. When $J$ is larger than 3.5, the d-wave pairing on $d_{xz}$ and $d_{yz}$ orbitals becomes stronger. The SC gap on the Fermi surfaces (FSs) are shown in Fig. 2(b) and (c) when $J=2.0$ and $J=4.0$. There are SC gap nodes along the $(\pi,\pi)$ direction.

\section{random phase approximation analysis}\label{sectioniii}
In this section, we perform calculations under the RPA approximation to obtain the pairing symmetry under the standard spin-fluctuation mechanism. The full Hamiltonian used in the following calculation is
\begin{equation}
\begin{aligned}
\emph{H}=&\emph{H}_{0}+U\sum_{i,\alpha}n_{i\alpha\uparrow}n_{i\alpha\downarrow}+U^{'}\sum_{i,\alpha<\beta}n_{i\alpha}n_{i\beta}\\
         &+J_{H}\sum_{i,\alpha<\beta,\sigma\sigma^{'}}c^{\dagger}_{i\alpha\sigma}c^{\dagger}_{i\beta\sigma^{'}}c_{i\alpha\sigma^{'}}c_{i\beta\sigma}\\
         &+J^{'}\sum_{i,\alpha\neq\beta}c^{\dagger}_{i\alpha\uparrow}c^{\dagger}_{i\alpha\downarrow}c_{i\beta\downarrow}c_{i\beta\uparrow},
\end{aligned}
\label{hh}
\end{equation}
where $H_0$ is the above tight binding three bands model
and $n_{i,\alpha}=n_{i,\alpha,\uparrow}+n_{i,\alpha,\downarrow}$. For other indexes, we adopt the parameter notations given in ref. \ucite{Kemper}. In the RPA approximation, the pairing vertex is
\begin{equation}
\begin{aligned}
\Gamma_{ij}(k,k^{'})=&\emph{Re}\bigg[\sum_{l_{1}l_{2}l_{3}l_{4}}a^{l_{2},\ast}_{\emph{v}_{i}}(k)a^{l_{3},\ast}_{\emph{v}_{i}}(-k)\\
&\times\Gamma_{l_{1}l_{2}l_{3}l_{4}}(k,k^{'},\omega=0)a^{l_{1}}_{\emph{v}_{j}}(k^{'})a^{l_{4}}_{\emph{v}_{j}}(-k^{'})\bigg],
\end{aligned}
\end{equation}
where the momenta $k$ and $k^{'}$ is restricted to different FSs with $k\in C_{i}$ and $k^{'}\in C_{j}$. $a^{l}_{v}$(orbital index $l$ and band index $v$) is the component of the eigenvectors of the three-orbital tight binding Hamiltonian.
The singlet channel of orbital vertex function $\Gamma_{l_{1}l_{2}l_{3}l_{4}}$ in RPA is given by
\begin{equation}
\begin{aligned}
\Gamma_{l_{1}l_{2}l_{3}l_{4}}(k,k^{'},\omega)=&\bigg[\frac{3}{2}\bar{U}^{s}\chi^{RPA}_{1}(k-k^{'},\omega)\bar{U}^{s}+\frac{1}{2}\bar{U}^{s}\\
&-\frac{1}{2}\bar{U}^{c}\chi^{RPA}_{0}(k-k^{'},\omega)\bar{U}^{c}+\frac{1}{2}\bar{U}^{c}\bigg]_{l_{3}l_{4}l_{1}l_{2}},
\end{aligned}
\end{equation}
where $\chi^{RPA}_{1}$ and $\chi^{RPA}_{0}$ are the spin and charge fluctuation RPA susceptibility, respectively. The spin and charge interaction matrix ($\bar{U}^{s}$, $\bar{U}^{c}$) are the same as in ref. \ucite{Kemper}.
The pairing strength function is
\begin{equation}
\begin{aligned}
\lambda\big[\emph{g}(k)\big]=-\frac{\sum_{ij}\oint_{C_{i}}\frac{d\emph{k}_{\|}}{\emph{v}_{\emph{F}}(k)}\oint_{C_{j}}\frac{d\emph{k}^{'}_{\|}}{\emph{v}_{\emph{F}}(k^{'})}\emph{g}(k)\Gamma_{ij}(k,k^{'})\emph{g}(k^{'})}{(2\pi)^{2}\sum_{i}\oint_{C_{i}}\frac{d\emph{k}_{\|}}{\emph{v}_{\emph{F}}(k)}\big[\emph{g}(k)\big]^{2}},
\end{aligned}
\end{equation}
where $v_{F}(k)=|\nabla_{k}E_{i}(k)|$ is the Fermi velocity on a given Fermi surface sheet $C_{i}$.
We perform calculations in the spin-rotational invariance case meaning $U^{'}=U-2J_{H}$ and $J_{H}=J^{'}$.

\begin{figure}
\centerline{\includegraphics[width=0.5\textwidth]{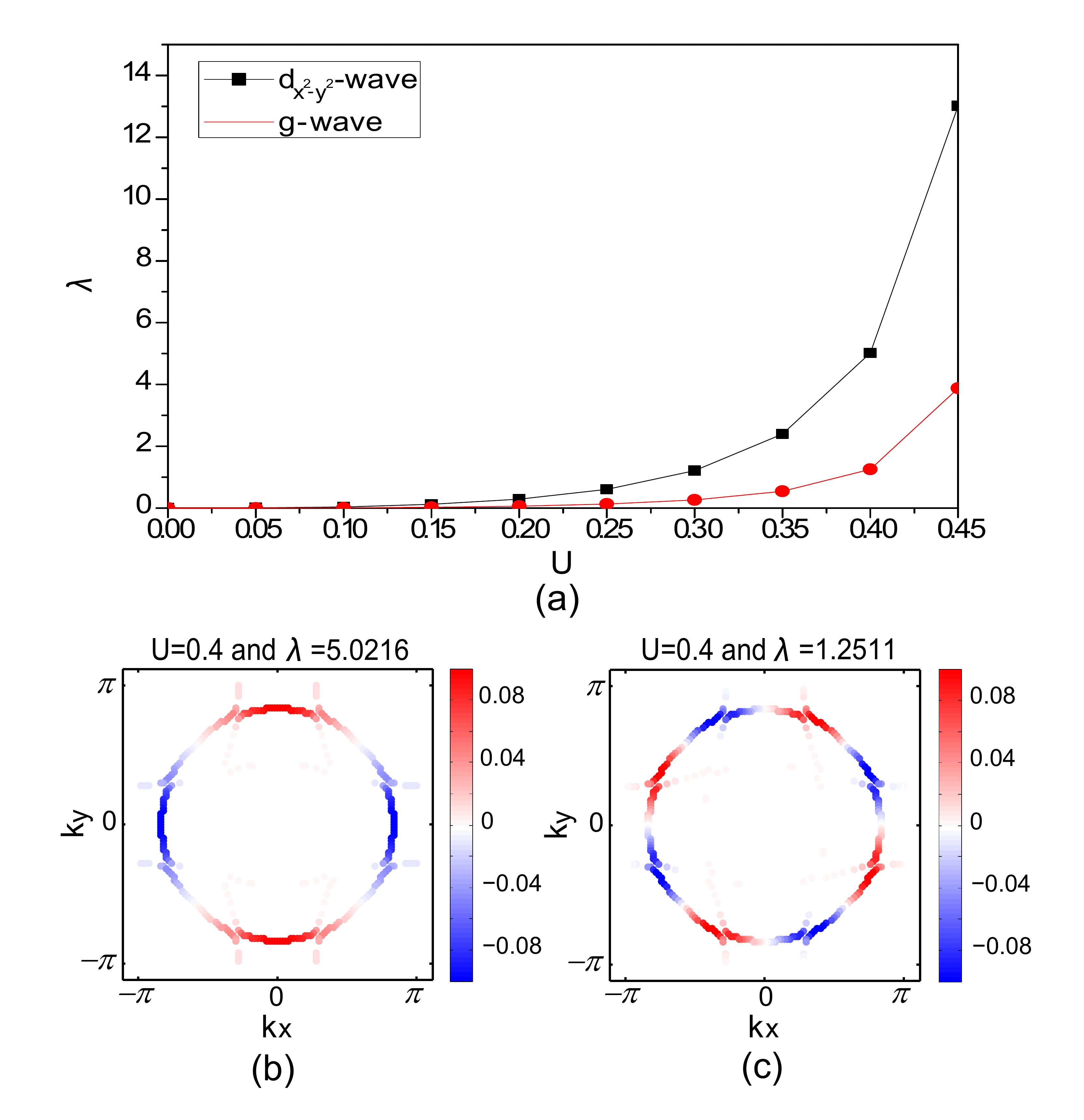}}
\caption{(color online) The pairing strengths and gap functions for $J_{H}/U=0.2$. The two largest pairing strengths as a function of $U$ in (a). The two dominant gap functions for $d_{x^{2}-y^{2}}$-wave(b) and $g$-wave(c) at $U=0.4$.}
\end{figure}

The bare susceptibility and RPA spin susceptibility are shown in Fig. 1(b). The strongest peak is at $(\pi,0.96\pi)$ which is slightly away from $Q=(\pi,\pi)$. Fig. 3(a) shows the two leading eigenvalues as a function of $U$ at $J_{H}/U=0.2$. The dominant pairing symmetry is the $d_{x^{2}-y^{2}}$ symmetry  with  gap nodes along the $(\pi,\pi)$ direction. Moreover, the pairing on the $d_{x^{2}-y^{2}}$ orbital is much  stronger than those on the $d_{xz/yz}$, as shown in Fig. 3(b) at $U=0.4$. Fig. 3(c) shows the gap function for the subdominant $g$-wave pairing on the Fermi surfaces. We also perform the calculation on the intra-pocket and inter-pocket pairing strength in the $d_{x^{2}-y^{2}}$-wave pairing. The intra-pocket pairing makes the major contribution. Comparing the gap function on the FS in Fig. 3(b) with the orbital contribution on the FS shown in Fig. 1(a), we can  conclude  that  the intra-pocket pairing is mainly from the $d_{x^{2}-y^{2}}$ orbital. The real space Fourier transform of the irreducible singlet vertex $\Gamma_{l_{1}l_{2}l_{3}l_{4}}(r-r^{'})$, also indicates that the intra-orbital pairing vertex of the $d_{x^{2}-y^{2}}$ orbital has the largest negative value on the NN bonds. This means that the $d_{x^{2}-y^{2}}$ orbital has the largest attractive interaction on the NN bonds. From these analysis on both the momentum space and real space, the RPA suggests that the $d_{x^{2}-y^{2}}$ orbital plays a dominant role in the process of pairing. These results are consistent with the mean field calculation with a small $J$ value in the previous section.

\section{FRG ANALYSIS}\label{sectioniv}

In this section, a FRG analysis is  preformed to analyze the possible SC phase for the Hamiltonian in Eq. \ref{hh} . The FRG method has been applied to obtain pairing symmetries in both cuprates and iron-based superconductors \cite{zhaiwang,Yuan}.  It is a very powerful method to analyze the various competing order tendencies in a system.

\begin{figure}
\centerline{\includegraphics[width=0.5\textwidth]{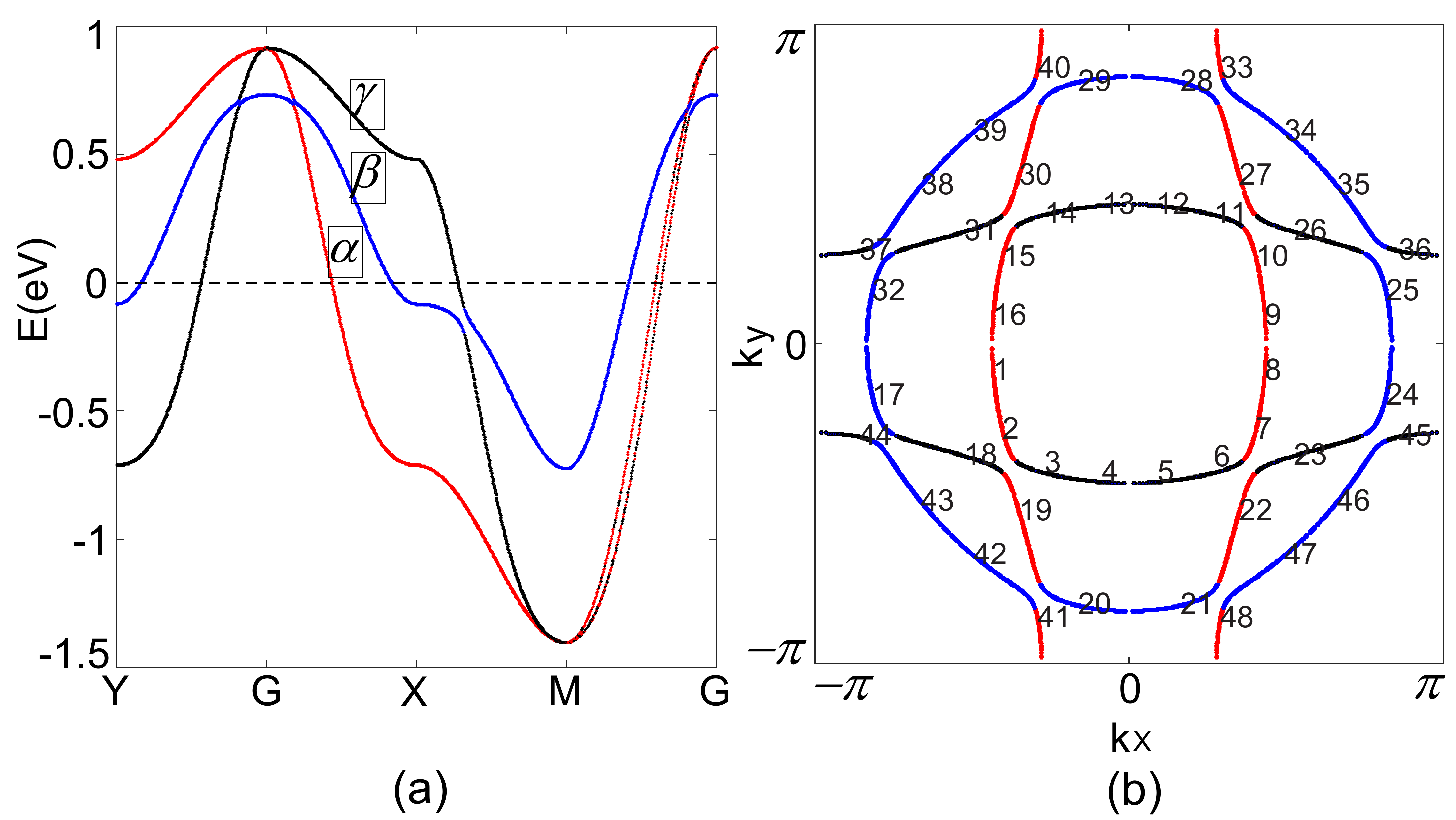}}
\caption{(color online) The division of the Fermi surfaces used in the FRG calculations with N=48.}
\label{fig:bdstr}
\end{figure}

The detailed description of the method can be found in ref. \ucite{Salmhofer,Metzner,Platt}.  Here we take the band basis and make the Fourier transformation to rewrite the interaction in the momentum space.  The interaction part at different flowing parameter $\Lambda$ is given by
\bea
H_{int}^{\Lambda}=\sum_{{\bf k}_1,...,{\bf k}_4}\sum_{n_1,...,n_4}\{&&V^{B}_{\Lambda}(n_1,{\bf k}_1;n_2,{\bf k}_2;n_4,{\bf k}_4;n_3,{\bf k}_3)\nn \\
\times &&c^{\dagger}_{n_1;{\bf k}_1}c^{\dagger}_{n_2;{\bf k}_2}c_{n_4;{\bf k}_4}c_{n_3;{\bf k}_3}\}
\eea
where $n_1$ to $n_4$ are the band index and $V^{B}_{\Lambda}$ is the interaction strength. Different competing order tendencies can be characterized by the instability of the corresponding channels. Taking the superconductor order parameter $\hat{O}_{\bf k}=c_{-{\bf k},\downarrow}c_{\bf k,\uparrow}$ as  an example, we can write it as $\sum_{{\bf k},{\bf p}}V^{SC}_{\Lambda}({\bf k},{\bf p})\hat{O}^{\dagger}_{\bf k}\hat{O}_{\bf p}$.  $V^{SC}$ is  decomposed into $ V^{SC}_{\Lambda}({\bf k},{\bf p})=\sum_{i} w_{i}(\Lambda) f^{*}_{i}(k)f_{i}(p)$ \cite{zhaiwang,Thomale}. The leading instability corresponds to the minimum $w_{i}(\Lambda)$.  Its symmetry information  is  provided by $f_i(p)$.

\begin{table}
\begin{tabular}{cccc|cccc}
\hline
\hline
 &  &$U=2eV$  & & & & $U=4eV$ &  \\
\hline
(a,b) & $\alpha$ & $\beta$ & $\gamma$ & (a,b) &$\alpha$  & $\beta$ & $\gamma$  \\
\hline
$\alpha$  & -0.00&-4.96 & -0.59 & $\alpha$  & -0.00 & -11.49 & -1.21 \\
 $\beta$   & -4.96 & -168.48 & -3.93 & $\beta$   & -11.49 & -146.29 & -7.24 \\
  $\gamma$  & -0.59 & -3.93 & -5.93 & $\gamma$   & -1.21 & -7.24 & -3.62 \\
  \hline
  \hline
\end{tabular}
\caption{The leading eigenvalues of scattering vertex of superconducting channel between intra and inter-pockets in $(U,J_H/U)=(2,0.2)$ and $(4,0.2)$ respectively near half filling. }
\label{tab:superconductivity}
\end{table}

\begin{figure}
\centerline{\includegraphics[width=0.5\textwidth]{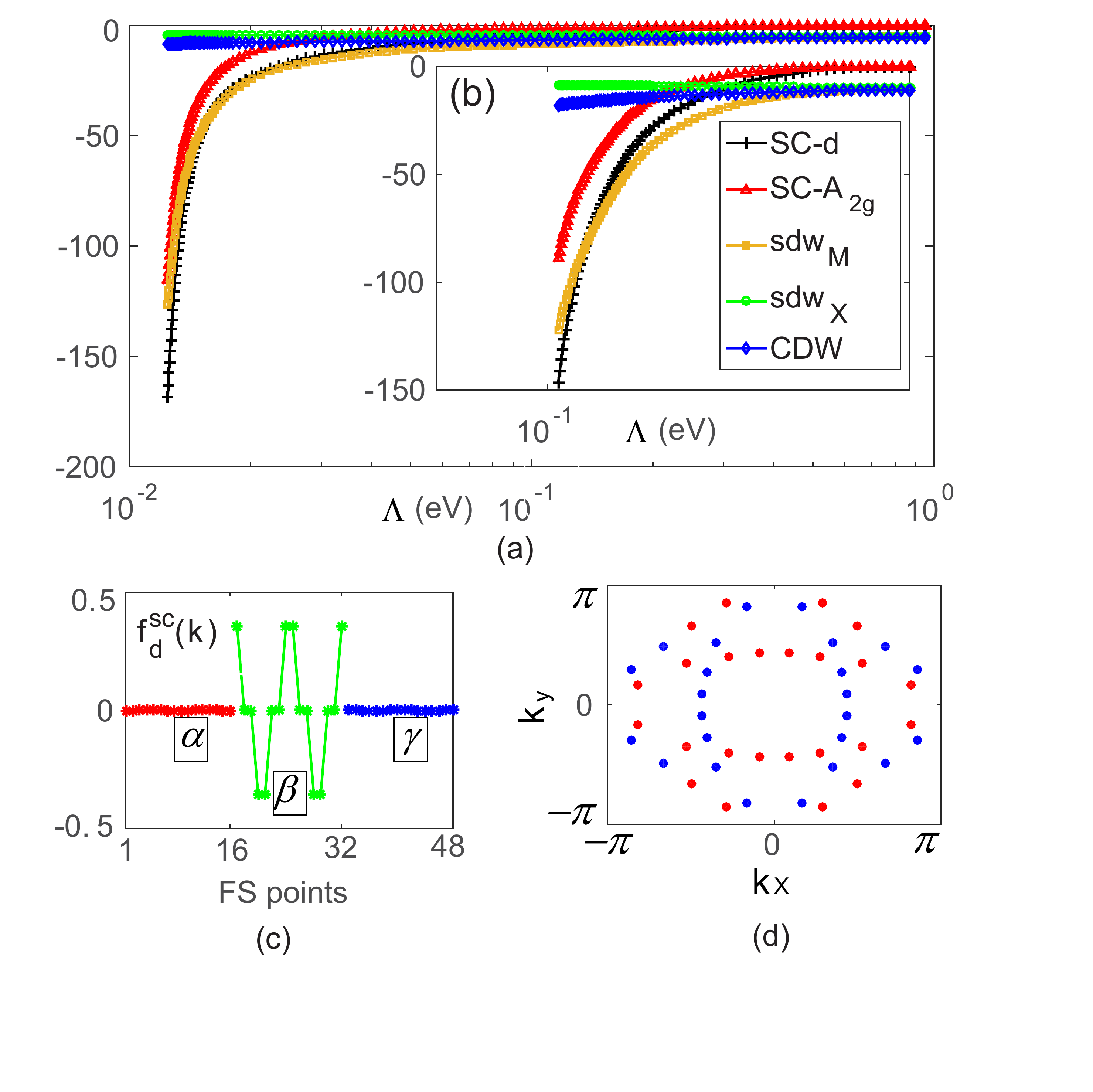}}
\caption{(color online) The FRG flow and  the   form factor of the superconducting order. (a-b):  The flows of different competing orders in the $(U,\frac{J_H}{U})=(2,0.2)$ and $(4,0.2)$ near half filling. The leading instability in both graphs is  the $d$-wave superconducting order. sdw$_{X}$ and sdw$_{M}$ is the spin density wave with propagating ${\bf Q}$ vector $(0,\pi)$ and $(\pi,\pi)$ respectively.  (c): The form factors of leading instability of  the superconducting order in the momentum space in the $(U,\frac{J_H}{U})=(2,0.2)$.  (d) The sign distribution of the d-wave superconducting order on the Fermi surfaces in which the red and blue filled circles represent the positive and negative sign of the pairing order respectively.}
\label{fig:flowsnearfilling}
\end{figure}

We divide the FS into N=48 patches as shown in Fig. \ref{fig:bdstr}(b). In Fig. \ref{fig:flowsnearfilling}, we report the FRG results near half filling with  $\frac{J_H}{U}=0.2$ and $U$ is chosen as $2eV$ and $4eV$ respectively.  The leading instability is a $d$ wave SC order. The pairing gap function $f({\bf k})$ as shown by Fig. \ref{fig:flowsnearfilling} has the largest weight in the $\beta$ band, which suggests that the pairing is orbital selective with dominant pairing on $d_{x^2-y^2}$ orbital.
This result is consistent with the RPA calculation. FRG methods are highly sensitive to the density state on FS. The $\beta$ band hosts a von Hove singularity near half filling while the other two bands, $\alpha$ and $\gamma$, have  large Fermi velocity near the Fermi level.  All these factors result in a dominant pairing in the $\beta$ band.

 We can further check the pairing vertex $V^{SC}(k,p)$ at the final stage.  We can compare the values of  the elements of the vertex.  These  elements with large values in this case exactly distribute on the $\beta$ band. We can also compare the intra and inter pocket contribution to pairing.   The scattering vertex of SC channel between intra and inter-pockets can be specified as $V_{a\rightarrow b}^{SC}(b,{\bf k};a,{\bf p})=V_f^{B}(b,{\bf k};b,-{\bf k};a,{\bf p};a,-{\bf p})$ where the symbol $a$ is the pocket index and $f$ means the final stage calculations. Decomposing these scattering vertex into different form factors, we list the leading eigenvalues  in Tab.\ref{tab:superconductivity}.   The scattering is dominated by the scattering within the $\beta$ pocket. This $d_{x^2-y^2}$ orbital selective pairing is very similar to the case in $Sr_2RuO_4$ \cite{Sr2RuO4} in which the  pairing is mainly from $d_{xy}$ \cite{Sr-triplet,QH} in FRG calculations. Interestingly, similar type of orbital selection occurs and is indicated explicitly on possible superconductivity in $LaOCrAs$\cite{QHWang}. The form factor $f({\bf k})$ in the $\beta$ pocket can be factorized to $A({\bf k})(\cos(k_x)-\cos(k_y))$ where the $A({\bf k})$ is the momentum-dependent amplitude, and it shares the same symmetry as the $d_{x^2-y^2}$ wave.   The $A_{2g}$-wave which changes sign under the mirror operation about the high symmetry lines is close to the $d$-wave when $U=2eV$ while it fades away slightly when $U=4eV$. This model also reveals the strong spin density wave (SDW) tendency in $(\pi,\pi)$ direction labeled by sdw$_{M}$ as illustrated in Fig. \ref{fig:flowsnearfilling}(a-b).

\begin{table}
\addtolength{\tabcolsep}{15pt}
\begin{tabular}{c|ccc}
\hline
\hline
 (a,b) &$\alpha$  & $\beta$ & $\gamma$  \\
\hline
 $\alpha$  & -2.44 & -2.20 & -2.04 \\
 $\beta$   & -2.20 & -4.01 & -7.87 \\
  $\gamma$   & -2.04 & -7.87 & -114.05 \\
  \hline
  \hline
\end{tabular}
\caption{The leading eigenvalues of scattering vertex of sdw$_M$ between intra and inter-pockets in electron doping  $1$ when $(U,J_H/U)=(3,0.1)$.}
\label{tab:sdw}
\end{table}

\begin{figure}
\centerline{\includegraphics[width=0.5\textwidth]{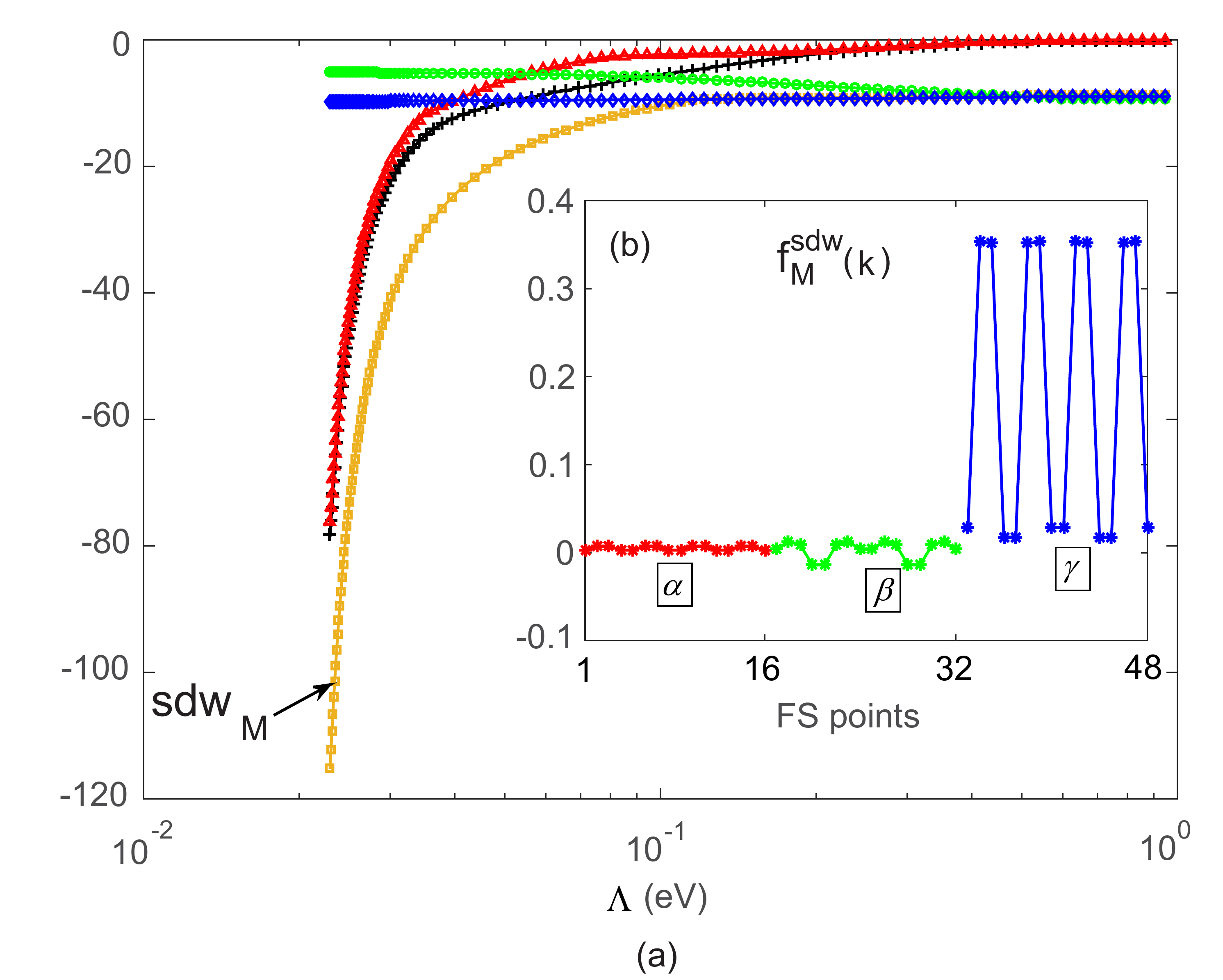}}
\caption{(color online) (a) The flows of different competing order in $(U,\frac{J_H}{U})=(3,0.1)$ when $\delta$ is $1$. The leading instability is the sdw$_M$ state. (b) The form factor of the sdw$_M$ state.  }
\label{fig:sdw}
\end{figure}


 The FRG result in the heavy electron doping region is reported in Fig. \ref{fig:sdw}, which is corresponding to near a doping level with one additional electron per site from the half filling. Namely, it is about one third electron doping for each $t_{2g}$ orbital. The interaction parameter here we use is $(U, \frac{J_H}{U})=(3,0.1)$. As manifested by Fig. \ref{fig:sdw}(a), the SC channel is not the leading instability any more, and the SDW channel with wavevector $(\pi,\pi)$ (sdw$_M$) arises. The sdw$_M$ stems from the nesting  on Fermi surfaces between points $43$ to $34$ as shown in Fig. \ref{fig:bdstr}. The form factor $f^{sdw}_M({\bf k})$ of $sdw_M$ is illustrated in Fig. \ref{fig:sdw}(b). The weights of the form factor are located at the $\gamma$ pocket. More precisely, the largest weights are distributed at the ${\bf k}$ points in the $\gamma$ pocket along the diagonal lines. In our Fermi surface patches, these points are located at $34-35$ and their counterparts under the $C_4$ symmetry. We can define the scattering vertex of sdw$_M$ between pockets as $V^{sdw_M}_{a \rightarrow b}(b,k;a,p)$ and decompoe it into different form factors as before. The leading eigenvalues are given in the Tab.\ref{tab:sdw}. From the Tab.\ref{tab:sdw},  the scatterings  in the sdw$_M$ channel are attributed to the $\gamma$ pocket and the contributions from others are very small.


\section{ Discussion and Conclusion}\label{sectionv}
In summary, we have shown that the $d$-wave superconducting state is a robust superconducting state in this family of materials. If we treat this system in a weak interaction region, the superconducting pairing is dominated by the intraorbital pairing of the $d_{x^2-y^2}$ orbital. The $d_{xz,yz}$ orbitals have very small contribution. However, in the strong correlation region, as the bare electron bands are strongly renormalized, all three $t_{2g}$ orbitals can have strong superconducting pairing.

These results suggest that the new materials can serve a bridge to connect cuprates and iron-based superconductors to solve unconventional high $T_c$ mechanism and  can be a ground to test the validity of theoretical methods.  In cuprates, the electronic physics is governed by  the single $e_g$ $d_{x^2-y^2}$ orbital. The d-wave pairing symmetry was obtained by all methods based on repulsive interactions \cite{Scalapino,FCZhang,Kotliar}. Thus it is difficult to distinguish  different theoretical methods.  In the new materials, the $t_{2g}$ $d_{x^2-y^2}$ essentially plays the  same role as the  $d_{x^2-y^2}$ orbital in cuprates. However, the difference among three $t_{2g}$ orbitals can directly distinguish different methods based on strong and weak correlations. Both FRG and RPA, which essentially are only valid in weak interaction region,  consistently predict an strongly orbital selective superconducting state. Thus measuring the superconducting gaps on the FSs attributed to the $d_{xz,yz}$ orbitals can determine the validity of these conventional theoretical approaches.

The superconducting pairing mechanism in iron-based superconductors has also become very controversial\cite{hu&yuan,chubukov,wanglee,mazin1,mazin2,Hirschfeld}. These new materials have the same multi-orbital electronic physics locally as iron-based superconductors. Because of the difference on their Fermi surface topologies near half filling,  the comparison between these two systems can help us to establish   general principles in determining superconducting pairing symmetry and pairing mechanisms.

\section{Acknowledgements}

 The work is supported by the Ministry of Science and Technology of China 973 program (No. 2015CB921300), National Science Foundation of China (Grant No. NSFC-1190020, 11534014, 11334012), and the Strategic Priority Research Program of CAS (Grant No.XDB07000000). QHW also acknowledges the supports by the NSFC funding No.11574134.

\end{document}